\documentclass[12pt]{iopart}
\expandafter\let\csname equation*\endcsname\relax
\expandafter\let\csname endequation*\endcsname\relax
\usepackage{mathtools} 
\usepackage{makeidx}
\usepackage{eurosym}
\usepackage{graphicx}
\usepackage{bm}
\usepackage{fancyhdr}
\usepackage{amsmath}
\usepackage{amsfonts}
\usepackage{amsbsy}
\usepackage{amssymb}
\usepackage[mathscr]{eucal}
\usepackage{multirow}
\usepackage{mciteplus}
\usepackage{color}

\newcommand{\beq}{\begin{equation}}
\newcommand{\eeq}{\end{equation}}
\newcommand{\ba}{\begin{array}}
\newcommand{\ea}{\end{array}}
\newcommand{\bea}{\begin{eqnarray}}
\newcommand{\eea}{\end{eqnarray}}
\newcommand{\bseq}{\begin{subequations}}
\newcommand{\eseq}{\end{subequations}}

\begin{document}

\title[Time averaging of weak values]{Time averaging of weak values -
consequences for time-energy and coordinate-momentum uncertainty.}
\author{Eli Pollak$^1$ and Salvador Miret-Art{\'e}s$^2$}
\address{$ˆ1$ Chemical and Biological Physics Department, Weizmann
Institute of Science, 76100 Rehovoth, Israel}
\address{$ˆ2$ Instituto de F\'isica Fundamental, Consejo Superior de Investigaciones
Cient\'ificas, Serrano 123, 28006 Madrid, Spain}
\eads{\mailto{$^1$
eli.pollak@weizmann.ac.il}, \mailto{$^2$ s.miret@iff.csic.es}}

\begin{abstract}
Using the quantum transition path time probability distribution we show that
time averaging of weak values leads to unexpected results. We prove a weak
value time energy uncertainty principle and time energy commutation
relation. We also find that time averaging allows one to predict in advance
the momentum of a particle at a post selected point in space with accuracy
greater than the limit of $\hbar /2$ as dictated by the uncertainty
principle. This comes at a cost - it is not possible at the same time to
predict when the particle will arrive at the post selected point. A specific
example is provided for one dimensional scattering from a square barrier.
\end{abstract}



\submitto{\NJP}

\maketitle


\section{Introduction}

The Heisenberg uncertainty principle \cite{heisenberg27,ballentine98} is a
cornerstone of quantum theory. It establishes a lower bound on the product
of the variances of non-commuting observables. More specifically, let $\hat{q%
}$ and $\hat{p}$ denote the position and momentum operators and $|\Psi
_{t}\rangle =\exp \left( -\frac{i\hat{H}t}{\hbar }\right) |\Psi _{0}\rangle $
a wavefunction evolved to time $t$ from its form $|\Psi _{0}\rangle $ at
time $0$ under the Hamiltonian operator $\hat{H}$. The mean values of the
position and the momentum at any time $t$ are $\bar{q}_{t}=\left\langle \Psi
_{t}\left\vert \hat{q}\right\vert \Psi _{t}\right\rangle ,\bar{p}%
_{t}=\left\langle \Psi _{t}\left\vert \hat{p}\right\vert \Psi
_{t}\right\rangle $. The Heisenberg uncertainty principle then assures us
that at any time $t$ the product of the variances $\Delta
q_{t}^{2}=\left\langle \Psi _{t}\left\vert \hat{q}^{2}-\bar{q}%
_{t}^{2}\right\vert \Psi _{t}\right\rangle $ and $\Delta
p_{t}^{2}=\left\langle \Psi _{t}\left\vert \hat{p}^{2}-\bar{p}%
_{t}^{2}\right\vert \Psi _{t}\right\rangle $ is bounded from below - $\Delta
q_{t}^{2}\Delta p_{t}^{2}\geq \hbar ^{2}/4$. In these relations time is
considered to be the external time as defined for example by Busch \cite%
{busch08}. In his words, it is "the time period between the preparation and
the instant at which a measurement of, say, position is performed." The
concept of time which we will consistently use in this paper is this
external time.

The experimental implications of the uncertainty relation are well
understood. Repeated independent measurement of the momentum and position of
particles, lead to an uncertainty which is greater or equal to the
Heisenberg lower bound \cite{ballentine98}. More specifically, equal copies
of the same particle are prepared at some initial time $t=0$. These are
described by the wavefunction $|\Psi _{0}\rangle $. The particles are
allowed to evolve up to some time $t$ and at this time, either their
position or their momentum is measured. Multiple repetition of such a
measurement will lead to the observation of a mean momentum and position and
standard deviations from the means whose product is bounded from below by $%
\hbar /2$. The uncertainty relation as presented above differs from the more
general measurement-disturbance relation of Ozawa \cite{ozawa03} which was
subsequently derived using weak values \cite{lund10} and verified
experimentally by Steinberg and coworkers \cite{rozema12}.

It should be emphasized that the constraint on the standard deviations holds
only if measurements are carried out after the same time interval $t$ passed
between preparation ($t=0$) and measurement. If one measures the momentum at
time $t_{1}$ and the position at time $t_{2}\neq t_{1}$ then the uncertainty
product relation for the standard deviations no longer holds in the simple
form presented above. It is important to note that the Heisenberg principle
does not prevent one from determining with high precision either the
coordinate or the momentum at a specified time $t$. It only disallows the
precise simultaneous determination of the values of both the coordinate and
the momentum at the same time $t$.

A central result of this paper is an analog of these last statements in
connection with a post selected value of the position $x$, analogous to the
precisely determined external time interval $t$ and determination of the
momentum. We will show that the Heisenberg relation does not prevent us from
constructing a scenario in which we can predict with a certainty which is
greater than the Heisenberg limitation, the momentum of a particle when it
reaches a post selected position $x$. The scenario we present precludes the
ability to predict accurately the time at which the particle arrives at the
point $x$. Different particles may arrive at $x$ \textit{at different times}
and therefore {\ the product of} their uncertainties in position and
momentum are not limited by the Heisenberg relations. We will {\ also} show
that the ability to {\ pinpoint} the momentum and position of the particle
comes at the price of not being able to predict at what time the particle
will reach the point $x$, if at all. We do know that if it reaches the point
$x$ it will have with some certainty a known momentum $p$.

To justify this assertion, it is necessary to consider the energy time
uncertainty relation. One of the challenges is that the simple derivation of
an uncertainty relation for the coordinate and momentum does not exist for
energy and time. Busch argues \cite{busch08} that "there is no unique
universal relation that could stand on equal footing with the
position-momentum uncertainty relation". With respect to the external time
he notes that "External time is sharply defined at all scales relevant to a
given experiment. Hence there is no scope for an uncertainty interpretation
with respect to external time." Hilgevoord \cite{hilgevoord96} claimed that
"there is no reason why a Heisenberg relation should hold ... between the
time coordinate and the energy of the system."

As originally discussed by Pauli \cite{pauli33} one of the difficulties in
defining energy time uncertainties comes from the fact that the energy
spectrum of the Hamiltonian operator is typically bounded from below. These
assertions notwithstanding, a second central result of this paper is the
derivation of a general energy - external time uncertainty principle based
on time averaging of weak values. To prevent any
misunderstanding, the energy time uncertainty relation we derive does not
involve a time operator. It cannot be over-stressed that in this paper, time
is considered only as an external time as defined by Busch, yet it is
measurable, as evidenced by the time dependent Schr{\"o}dinger equation.

The results described above will be derived by using a combination of two
formalisms. One is the quantum transition path time formalism \cite%
{pollak17c}, by which one determines the (external) time distribution of a
particle localized initially ($t=0$) as described by the wavefunction $\Psi
_{0}$, reaching the point $x$ at time $t$. The second formalism to be used
is that of weak values \cite{aharonov88,wiseman07,mitchison07,tamir13}. The
new element to be introduced in this paper is the combination of the two
formalisms, that is time averaging of weak values using the transition path
time distribution. The results derived in this paper are based on this time
averaging of weak values. Our assertions will be supported with some model
computations for the tunneling scattering of a particle through a one
dimensional square barrier. These model computations exemplify
the practical implications of the weak value time energy uncertainty relation stressing
again, that there is no need to define a time operator. All that is needed
are time of flight measurements. 

\section{Time averaging and a weak value energy time uncertainty principle}

Limiting the discussion to one dimension, we use the definition of the normalized transition
path time distribution for finding a particle at the post selected point $x$ as in Ref. \cite{petersen17}:
\begin{equation}
P\left( t;x\right) =\frac{\left\langle \Psi _{t}\left\vert \delta \left(
\hat{q}-x\right) \right\vert \Psi _{t}\right\rangle }{\int_{0}^{\infty
}dt\left\langle \Psi _{t}\left\vert \delta \left( \hat{q}-x\right)
\right\vert \Psi _{t}\right\rangle }\equiv \frac{\left\vert \left\langle
x|\Psi _{t}\right\rangle \right\vert ^{2}}{N\left( x\right) }.  \label{1}
\end{equation}%
It gives the probability of finding the system at time $t$ at the post
selected point $x$. Implicit in this definition is the assumption that the
normalization time integral in the denominator converges.

We note here at the outset that the transition path time
distribution is in principle measurable using a suitably defined time of
flight experiment. One places a screen at the post-selected point $x$ and
two synchronized clocks, one at the orifice of the emerging particles and
the other at the screen. For example, in single atom time-of-flight
experiments \cite{Fuhrmanek2010}, particles are released from a trap \cite%
{Du2015} at time zero, and the arrival time at a detector screen is
recorded. 

The weak value of an operator $\hat{O}$ at the post selected point x is
defined as \cite{tamir13}
\begin{equation}
O_{w}\left( x,t\right) =\frac{\left\langle x\left\vert \hat{O}\right\vert
\Psi _{t}\right\rangle }{\left\langle x|\Psi _{t}\right\rangle }.  \label{2}
\end{equation}%
It is well known that the spatial average of the weak value is
identical to the result of a strong measurement, that is:%
\begin{equation}
\int_{-\infty }^{\infty }dx\left\vert \left\langle x|\Psi _{t}\right\rangle
\right\vert ^{2}O_{w}\left( x,t\right) =\left\langle \Psi _{t}\left\vert
\hat{O}\right\vert \Psi _{t}\right\rangle .  \label{2a}
\end{equation}%
This does not mean that one cannot measure the weak value precisely, indeed,
repeated weak measurement experiments in which the weak value is measured
after the same time interval $t$ and at the same post-selected position $x$
will give the weak value as defined in Eq. \ref{2}. However, at different
post-selected values of the coordinate $x$, one will find different weak
values. Their spatial average as defined in Eq. \ref{2a} will lead to the
strong value.

Similarly, and this is a central new concept introduced in this paper, one
may define a time averaged mean of the weak value
\begin{equation}
\left\langle O\left( x\right) \right\rangle \equiv \int_{0}^{\infty
}dtP\left( t;x\right) O_{w}\left( x,t\right) .  \label{3}
\end{equation}%
Measuring the weak value at different times will give different
results. Time averaging them will give the time averaged weak value as
defined in Eq. \ref{3}. In different words, consider the time of flight
experiment. The screen is located at the post-selected point $x$. The
particle will arrive at the screen at different times. For each fixed time
there will be a weak value which may be measured. It will though be
different at different times. Its time average is defined in Eq. \ref{3}. In
the following we will use the notation%
\begin{equation}
\left\langle O_{1}\left( x\right) O_{2}\left( x\right) \right\rangle \equiv
\int_{0}^{\infty }dtP\left( t;x\right) O_{1,w}\left( x,t\right)
O_{2,w}\left( x,t\right)   \label{3a}
\end{equation}%
to denote the time average of a product of weak values. We will also use the bracket notation for moments of the time parameter:
\begin{equation}
\langle t^n(x)\rangle=\int_0^{\infty}dtt^nP(t;x)
\label{3b}
\end{equation}

With these preliminaries, following the standard derivation of the
uncertainty principle \cite{cohentannoudji77} we consider the inequality
\begin{equation}
0\leq \frac{1}{N\left( x\right) }\int_{0}^{\infty }dt\left\langle
x\left\vert t{\hat{I}}-i\lambda \hat{H}\right\vert \Psi _{t}\right\rangle
\left\langle \Psi _{t}\left\vert t{\hat{I}}+i\lambda \hat{H}\right\vert
x\right\rangle  \label{4}
\end{equation}%
where $\lambda $ is an arbitrary real number, $t$ is the scalar value of the
time, $\hat{I}$ is the identity operator and $\hat{H}$ is the Hamiltonian
operator. We stress that the time as used here is just a
parameter, not an operator. It multiplies the identity operator which is of
course hermitian. Therefore the product as defined on the r.h.s of Eq. \ref%
{4} is necessarily positive. Noting that
\begin{equation}
\hat{H}|\Psi _{t}\rangle =i\hbar \frac{\partial }{\partial t}|\Psi
_{t}\rangle  \label{5}
\end{equation}%
allows us to rewrite the inequality in \ref{4} as:
\begin{equation}
0\leq \int_{0}^{\infty }dtt^{2}P\left( t;x\right) +\lambda \hbar
\int_{0}^{\infty }dtt\frac{\partial }{\partial t}P\left( t;x\right) +\lambda
^{2}\int_{0}^{\infty }dtP\left( t;x\right) \frac{\left\langle x\left\vert
\hat{H}\right\vert \Psi _{t}\right\rangle \left\langle \Psi _{t}\left\vert
\hat{H}\right\vert x\right\rangle }{\left\langle x|\Psi _{t}\right\rangle
\left\langle \Psi _{t}|x\right\rangle }.  \label{6}
\end{equation}%
Due to the introduction of time averaging and the assumption that the
normalization integral $N\left( x\right) <\infty $, one may integrate the
middle term on the right hand side by parts to find:%
\begin{equation}
0\leq \left\langle t^{2}\left( x\right) \right\rangle -\lambda \hbar
+\lambda ^{2}\left\langle H\left( x\right) H^{\ast }\left( x\right)
\right\rangle .  \label{7}
\end{equation}%
Minimizing with respect to $\lambda $ leads to the time averaged weak value
second moment energy and time relation:%
\begin{equation}
\left\langle t^{2}\left( x\right) \right\rangle \left\langle H\left(
x\right) H^{\ast }\left( x\right) \right\rangle \geq \frac{\hbar ^{2}}{4}.
\label{8}
\end{equation}

Continuing in this vein, consider the relation between the standard
deviations. Denoting
\begin{equation}
\Delta t=t-\left\langle t\left( x\right) \right\rangle ,\Delta \hat{H}=\hat{H%
}-\left\langle H\left( x\right) \right\rangle  \label{9}
\end{equation}%
and using as before the inequality%
\begin{equation}
0 \leq \frac{1}{N\left( x\right) }\int_{0}^{\infty }dt \left\langle
x\left\vert \Delta t{\hat I}-i\lambda \Delta \hat{H}\right\vert \Psi
_{t}\right\rangle \left\langle \Psi _{t}\left\vert \Delta t{\hat I}+i\lambda
\Delta \hat{H}^{\dag}\right\vert x\right\rangle,  \label{10}
\end{equation}%
the relationship as in \ref{5} and integrating by parts one readily finds
\begin{equation}
0\leq \left\langle \Delta t^{2}\left( x\right) \right\rangle +\lambda
^{2}\left\langle \Delta \hat{H}\left( x\right) \Delta \hat{H}^{* }\left(
x\right) \right\rangle - \lambda \hbar \left[ 1-\left\langle t\left(
x\right) \right\rangle P\left( 0;x\right) \right] .  \label{11}
\end{equation}%
Minimizing with respect to $\lambda $ gives the central result of this
section, namely the uncertainty relation for the standard deviations:
\begin{eqnarray}
\sqrt{\left\langle \Delta t^{2}\left( x\right) \right\rangle \left\langle
\Delta \hat{H}\left( x\right) \Delta \hat{H}^{*}\left( x\right)
\right\rangle }\geq \frac{\hbar }{2}\left[ 1-\left\langle t\left( x\right)
\right\rangle P\left( 0;x\right) \right] .  \label{12}
\end{eqnarray}%
If the post selected coordinate $x$ is sufficiently far away from the
incident wavepacket then $P\left( 0;x\right) =0$ and we have regained a time
energy uncertainty relation for the time averaged weak values which is
identical to Heisenberg's result for coordinate momentum uncertainty.

{\ The term $\left\langle t\left( x\right) \right\rangle P\left( 0;x\right)$
appears in (\ref{12}) since we have imposed that the time measurement is in
the interval $[0,\infty]$ and not $[-\infty,\infty]$. In a typical
experimental setup, the particle cannot be initially found at the point $x$
so that effectively we have regained the lower limit of $\hbar/2$. 
This uncertainty relation implies that} if the standard deviation of the
time averaged weak value of the energy is small, then the uncertainty in the
time of arrival at the point $x$ becomes very large.

The time averaged weak value of the energy-time uncertainty relation is
intimately related to a time averaged weak value commutator of the energy
and the time defined by taking into consideration that the weak value of the
Hamiltonian operator is complex. One may define a weak time value
\begin{eqnarray}
t_w(x)=\frac{\langle x\vert t{\hat I}\vert\Psi_t\rangle}{\langle x\vert\Psi_t\rangle}=t.
\label{12aa}
\end{eqnarray}
To prevent misunderstanding, here too, the time is not considered as an operator, as before, it is the
external time and therefore this value of the time is just the time itself.  One then readily finds, using Eq. \ref{5} that%
\begin{eqnarray}
\left\langle \left[ H\left( x\right) ,t\left( x\right) \right] \right\rangle
&\equiv & \left\langle \left[ t\left( x\right)H^{*}\left( x\right) -H\left(
x\right) t^{*}\left( x\right) \right] \right\rangle  \notag \\
&=& \int_{0}^{\infty }dtP\left( t;x\right) t\left[ \frac{\left\langle \Psi
_{t}\left\vert \hat{H}\right\vert x\right\rangle }{\left\langle \Psi
_{t}|x\right\rangle }-\frac{\left\langle x\left\vert \hat{H}\right\vert \Psi
_{t}\right\rangle }{\left\langle x|\Psi _{t}\right\rangle }\right] =i\hbar
\label{12a}
\end{eqnarray}


\section{Predicting the momentum of a particle at a point x}

In this section we shall show that time averaging of weak values leads to
the conclusion that it is possible to predict the momentum of a particle at
a post selected value of the coordinate $x$ with arbitrary accuracy. For
this purpose we consider a scattering system, with a potential $V\left(
x\right) $ of finite range localized about $x=0$. The Hamiltonian of the
particle whose mass is $M$ is:%
\begin{equation}
\hat{H}=\frac{\hat{p}_{x}^{2}}{2M}+V\left( \hat{x}\right) .  \label{13}
\end{equation}%
To simplify we impose the condition that the potential function $V\left(
x\right) $ goes to $0$ as $x\rightarrow \pm \infty $ . Initially the system
is prepared in a coherent state $|\Psi _{0}\rangle $ localized about the
initial position $x_{i}$ chosen to be sufficiently far to the left of the potential such that $V\left( x_{i}\right) =0$, and
incident mean momentum $p_{i}>0$:%
\begin{equation}
\left\langle x|\Psi _{0}\right\rangle =\left( \frac{\Gamma }{\pi }\right)
^{1/4}\exp \left( -\frac{\Gamma \left( x-x_{i}\right) ^{2}}{2}+\frac{i}{%
\hbar }p_{i}\left( x-x_{i}\right) \right)  \label{14}
\end{equation}%
We will also assume that the probability of the particle initially leaking
into the interaction region ($x\sim 0$) is negligible ($\Gamma x_{i}^{2}\gg
1 $) where $\Gamma $ is the width parameter of the coherent state. This
implies that initially the particle is a free particle with positive mean
momentum $p_{i}$ in the $x$ direction. This initial state obeys the
Heisenberg position momentum uncertainty relation. The transition path time
distribution and the mean time of arrival at the post selected point are
well defined since for this generic scattering system it has been shown by
Muga \cite{muga08a} that at long time $\left\langle x|\Psi _{t}\right\rangle
\sim t^{-3/2}$. In the following we will choose the position $x > 0$ far
enough in the asymptotic region of the potential. Under such conditions, the
normalization integral $N\left( x\right) $ (see Eq. \ref{1}) becomes
independent of $x.$

The weak value of the momentum at the post-selected point $x$ is by
definition%
\begin{equation}
p_{w}\left( t;x\right) =\frac{\left\langle x\left\vert \hat{p}\right\vert
\Psi _{t}\right\rangle }{\left\langle x|\Psi _{t}\right\rangle }=-i\hbar
\frac{\partial \ln \left\langle x|\Psi _{t}\right\rangle }{\partial x}.
\label{15}
\end{equation}%
In this formulation, the position is known precisely, its post selected
value is $x$. It is well known \cite{tamir13} that the {\ spatial} average
of the weak value is the mean value of the momentum:
\begin{equation}
\left\langle \Psi _{t}\left\vert \hat{p}\right\vert \Psi _{t}\right\rangle
=\int_{-\infty }^{\infty }dx\left\vert \left\langle x|\Psi \left( t\right)
\right\rangle \right\vert ^{2}p_{w}\left( t;x\right) \text{.}  \label{16}
\end{equation}%
As discussed already in the previous section, instead of considering the
spatial average we will consider the time averaged weak value of the
momentum $\left\langle p\left( x\right) \right\rangle $ and its variance%
\begin{equation}
\left\langle \Delta p^{2}\right\rangle =\left\langle \left\vert p\left(
x\right) \right\vert ^{2}\right\rangle -\left\langle p\left( x\right)
\right\rangle ^{2}.  \label{17}
\end{equation}%
using the transition path time distribution and as before the brackets
denote time averages.

The weak value of the momentum is a complex quantity. Since we chose the
post selected value $x$ to be large enough such that the normalization
integral $N\left( x\right) $ is independent of $x$ we find that the
imaginary part of the weak value of the momentum is
\begin{equation}
\mathrm{Im}[p_{w}\left( t;x\right) ]=-\frac{\hbar }{2}\frac{\partial \ln
P\left( t;x\right) }{\partial x}  \label{18}
\end{equation}%
This means that the imaginary part of the time averaged value of the weak
momentum vanishes ($\mathrm{Im}{\left\langle p\left( x\right) \right\rangle
=0}$).

To obtain further insight into time averaged weak values, we consider
expressly the time evolved wavefunction, expanding it in terms of the
scattering eigenstates of the Hamiltonian:
\begin{equation}
\left\langle x|\Psi \left( t\right) \right\rangle =\int_{-\infty }^{\infty
}dp\exp \left( -i\frac{p^{2}t}{2M\hbar }\right) \left\langle
x|p^{+}\right\rangle \left\langle p^{+}|\Psi \right\rangle .  \label{19}
\end{equation}%
Asymptotically, the eigenfunctions $\left\langle x|p^{+}\right\rangle $ have
the form (with $p>0$):
\begin{equation}
\left\langle x|p^{+}\right\rangle =\left\{
\begin{array}{ll}
\frac{1}{\sqrt{2\pi \hbar }}\left[ \exp \left( \frac{ipx}{\hbar }\right)
+R\left( p\right) \exp \left( -\frac{ipx}{\hbar }\right) \right] , {%
x\rightarrow -\infty } &  \\
\frac{1}{\sqrt{2\pi \hbar }}T\left( p\right) \exp \left( \frac{ipx}{\hbar }%
\right) \ \ \ \ \ \ \ \ \ \ \ \ \ \ \ \ \ \ \ \ \ \ ,x\rightarrow \infty &
\end{array}%
\right.  \label{20}
\end{equation}%
and $R\left( p\right) ,T\left( p\right) $ are the reflection and
transmission amplitudes respectively. It is straightforward to evaluate the
time dependent wavefunction in the asymptotic region. Since the wavefunction
is initially localized outside of the range of the potential, the overlap $%
\left\langle p^{+}|\Psi \right\rangle =\left\langle p|\Psi \right\rangle
+R^{\ast }\left( p\right) \left\langle -p|\Psi \right\rangle $ where $%
\left\langle p|\Psi \right\rangle $ is just the momentum representation of
the initial wavefunction. Similarly, by choosing the post selected value of $%
x$ to be positive and large we have that $\left\langle x|p^{+}\right\rangle
=T\left( p\right) \left\langle x|p\right\rangle $. One is thus left with a
quadrature to obtain the time dependent wavefunction $\left\langle x|\Psi
\left( t\right) \right\rangle $ and its associated weak momentum value.

This quadrature needs to be carried out numerically, and is dependent on the
specifics of the potential which determines the momentum dependence of the
reflection and transmission amplitudes. However, one may readily obtain
analytic expressions using a steepest descent estimate of the integrals. One
finds that the important contribution to the time dependent wavefunction,
when $x\gg 0$ is%
\begin{eqnarray}
\left\langle x|\Psi \left( t\right) \right\rangle &\simeq & \left( \frac{%
\Gamma M^{2}}{\pi \left( M+it\hbar \Gamma \right) ^{2}}\right) ^{\frac{1}{4}%
}T\left( \frac{Mp_{i}-i\hbar \Gamma M\left( x_{i}-x\right) }{\left(
M+it\hbar \Gamma \right) }\right)  \notag \\
& \cdot & \exp \left( -\frac{p_{i}^{2}}{2\hbar ^{2}\Gamma }+\frac{M\Gamma }{2%
}\frac{\left( i\left( x_{i}-x\right) -\frac{p_{i}}{\hbar \Gamma }\right) ^{2}%
}{\left[ M+it\hbar \Gamma \right] }\right) .  \label{21}
\end{eqnarray}%
so that
\begin{equation}
\left\vert \left\langle x|\Psi \left( t\right) \right\rangle \right\vert
^{2} \simeq \frac{M\sqrt{\Gamma }}{\sqrt{\pi \left[ M^2+t^2\hbar^2 \Gamma^2 %
\right] }}\left\vert T\left( \frac{p_{i}-i\hbar \Gamma \left( x_{i}-x\right)
}{\left( 1+i\frac{t\hbar \Gamma }{M}\right) }\right) \right\vert ^{2} \exp
\left( -\frac{\Gamma \left( x_{i}-x+\frac{p_{i}t}{M}\right) ^{2}}{\left[
1+\left( \frac{t\hbar \Gamma }{M}\right) ^{2}\right] }\right).  \label{22}
\end{equation}%
The denominator of the transition path time distribution is estimated as:
\begin{equation}
\int_{0}^{\infty }dt\left\vert \left\langle x|\Psi \left( t\right)
\right\rangle \right\vert ^{2}\simeq M\left\vert T\left( p_{i}\right)
\right\vert ^{2}/p_{i}  \label{23}
\end{equation}%
and as noted, is independent of $x$. Within this steepest descent evaluation
the weak value of the momentum is:
\begin{equation}
p_{w}\left( t;x\right) \simeq \frac{M\left[ Mp_{i}-\hbar ^{2}\Gamma
^{2}\left( x_{i}-x\right) t\right] }{\left[ M^{2}+\left( t\hbar \Gamma
\right) ^{2}\right] } +i\frac{\hbar \Gamma M\left[ M\left( x-x_{i}\right)
-p_{i}t\right] }{\left[ M^{2}+\left( t\hbar \Gamma \right) ^{2}\right] }.
\label{24}
\end{equation}%
Time averaging this expression using the steepest descent estimate for the
transition path time distribution gives the result:
\begin{equation}
\left\langle p\left( x\right) \right\rangle \simeq p_{i}  \label{25}
\end{equation}%
or in other words, the time average of the weak value of the momentum equals
the initial averaged incident momentum. It remains to consider the second
moment of the weak value, and within the steepest descent approximation one
finds
\begin{equation}
\left\langle \Delta p^{2}\left( x\right) \right\rangle =\int_{0}^{\infty
}dtP\left( t;x\right) \left( \left\vert p_{w}\left( t;x\right) \right\vert
^{2}-\left\langle p(x)\right\rangle ^{2}\right) \simeq \frac{\hbar
^{2}\Gamma }{2}  \label{26}
\end{equation}%
which is precisely the momentum variance of the initial wavepacket. By
reducing the width parameter $\Gamma $ this variance can become arbitrarily
small.

We have thus demonstrated, using a steepest descent approximation that the
time average of the weak value of the momentum and its variance at the
post-selected value of the coordinate $x$ are the same as the initial mean
values $\bar{p}_{i}$ and $\Delta p_{i}^{2}$. However, the coordinate is
post-selected, it is known precisely. This means that the Heisenberg
relation does not limit the precision with which the time averaged post
selected weak value of the momentum and its variance may be determined.
Moreover, when $\Gamma $ is sufficiently small, the weak value of the
momentum will be very close to the incident mean value of the momentum. In
other words, even for a single particle, we can predict in advance its
momentum when reaching the post selected point $x$. This is a central result
of this paper.

The steepest descent approximation for the transition path time distribution
as given in Eq. \ref{22} goes at long times as $t^{-1}$ so that strictly
speaking the time integrals would diverge. As already mentioned, the correct
long time dependence of the transition path time distribution goes as $%
t^{-3} $ so that there is no problem in reality. The steepest descent
approximation is correct for finite time, the long time tail is very small,
it goes as $\exp\left( -p_{i}^{2}/\left[ \hbar ^{2}\Gamma \right] \right) $.
{\ To prevent any doubt, we} have also undertaken a numerically exact study
of a model system to demonstrate that indeed one may predict the momentum at
$x$ with arbitrary accuracy.

{\ We} consider scattering through a square barrier (atomic units are used
throughout) with a particle of mass $1/2$, a barrier height of unity,
barrier width $2$ and an incident momentum of $1/4$. The incident wavepacket
is chosen such that the initial variance of its momentum is small, the width
parameter of the coherent state is chosen to be $\Gamma =0.001$. We then
plot in Fig. 1 the transition path time probability distribution (\ref{1})
for $x=-x_{i}=100$ and compare it with its steepest descent approximation%
\begin{equation}
P_{SD}\left( t;x\right) = \frac{\sqrt{\Gamma }p_{i}}{\sqrt{\pi \left[
M^{2}+t^{2}\hbar ^{2}\Gamma ^{2}\right] }}\exp \left( -\frac{\Gamma \left(
x_{i}-x+\frac{p_{i}t}{M}\right) ^{2}}{\left[ 1+\left( \frac{t\hbar \Gamma }{M%
}\right) ^{2}\right] }\right).  \label{27}
\end{equation}

\begin{figure}[tbp]
\begin{centering}
\includegraphics[scale=0.4]{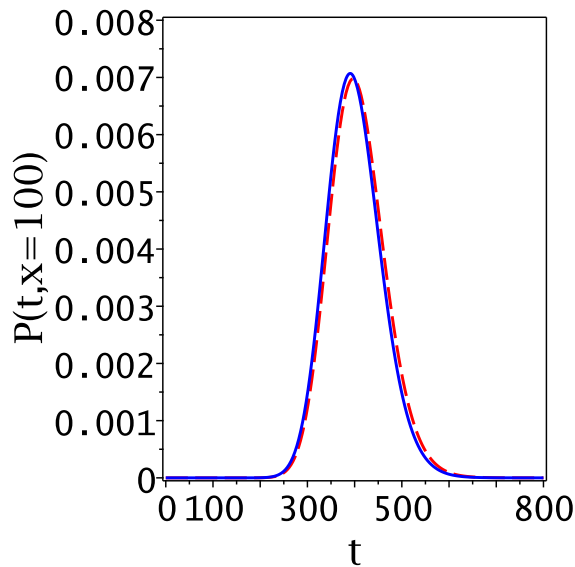} \bigskip
\par\end{centering}
\bigskip
\par
\bigskip
\caption{The transition path time distribution of a particle tunneling
through a square potential barrier. The solid (blue) line shows the
normalized transition path time distribution (Eq. \protect\ref{1}), the
dashed (red) line shows its Gaussian approximation ( \protect\ref{27}).
Details of the numerical values used are given in the text. }
\label{fig1}
\end{figure}
As is evident from the figure, the agreement is quantitative. The accurate
transition path time distribution {\ has its maximum at a time which is a
bit shorter} than the steepest descent approximation. Similarly, the value
of the time averaged weak momentum for these parameters is $0.2522$,
slightly higher than the steepest descent estimate ($1/4$) which does not
take into consideration the filtering effect of the transmission
probability. Due to the tunneling, the transmission favors higher momenta as
discussed in Ref. \cite{petersen17}. Decreasing the width parameter by a
factor of $4$, {\ reduces the value of the time averaged weak momentum to} $%
0.2505$.

The deviation of the real and imaginary parts of the weak value of the
momentum from their mean $\delta p_{w}\left( t;x\right) =p_{w}\left(
t;x\right) -\left\langle p\left( x\right) \right\rangle $ are plotted as a
function of time in Fig. 2 and compared with the steepest descent estimates
of Eq. \ref{24}. The early times lead to positive values of the momentum
differences, the later times to negative values, as might have been expected
from a classical mechanics perspective. The agreement between the steepest
descent estimates and the numerically exact estimates of the real and
imaginary parts of the weak values is quantitative.

Finally, the standard deviation ($\sqrt{\left\langle \Delta p^{2}\left(
x\right) \right\rangle }$) of the time averaged weak value of the momentum
with $\Gamma =0.001$ is found to be $0.02228$. From \ref{26} one finds the
value $0.02236$. The standard deviation is an order of magnitude less than
that of the mean value of the momentum itself. Reducing the width parameter
by a factor of $4$ gives a time averaged mean of the weak momentum of $%
0.2505 $ and a standard deviation of $0.01117$. As predicted in \ref{26} one
may arbitrarily reduce the standard deviation by reducing the width
parameter of the initial coherent state. In other words, in principle, using
time averaged weak values, one may accurately predict both the location and
the momentum of a single particle.

\begin{figure}[tbp]
\begin{centering}
\includegraphics[scale=0.4]{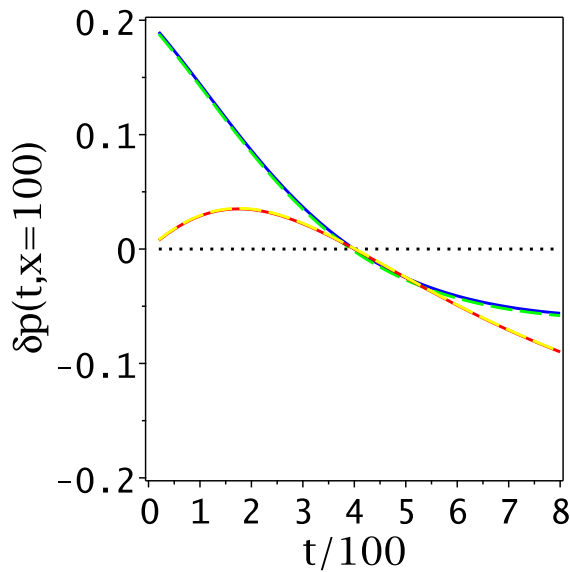} \bigskip
\par\end{centering}
\par
\bigskip
\par
\bigskip
\par
\bigskip
\caption{The time dependence of the deviation of the weak value of the
momentum from its mean at $x=-x_i=100$. The lower (red) dashed line shows
the numerically exact and solid (yellow) line shows the steepest descent
estimate for the deviation of the real part of the weak value from its mean.
The upper solid (blue) line shows the numerically exact and the dashed
(green) line the steepest descent estimate for the imaginary part of the
weak value of the momentum whose mean vanishes. The parameters used are
given in the text, the initial momentum of 0.25 is larger than the
deviation, which when weighted by the transition path time distribution (see
Fig. \protect\ref{fig1}) becomes rather small. }
\label{fig2}
\end{figure}

Does this imply that we may construct a quantum trajectory analogous to the
classical trajectory where the momentum and coordinate are known as
functions of the time? Of course not. Strictly speaking, the variance of the
mean time diverges logarithmically. Within the steepest descent estimate one
finds
\begin{equation}
\left\langle \Delta t^{2}\left( x\right) \right\rangle =\left\langle
t^{2}\left( x\right) \right\rangle -\left\langle t\left( x\right)
\right\rangle ^{2}\simeq \frac{1}{2\Gamma }\left( \frac{M}{p_{i}}\right) ^{2}
\label{28}
\end{equation}%
and this grows indefinitely as the precision with which the momentum is
predetermined increases, that is, as $\Gamma \rightarrow 0$. In other words,
the price paid for the precise determination of the momentum is imprecision
in the knowledge of when the particle will actually reach the post selected
point $x$. Within the steepest descent approximation one finds that the time
averaged weak value of the energy is
\begin{equation}
\left\langle H\left( x\right) \right\rangle \simeq \frac{p_{i}^{2}}{2M}+%
\frac{\hbar ^{2}\Gamma }{4M}  \label{29}
\end{equation}%
while the variance is given by:
\begin{equation}
\left\langle \left\vert H\left( x\right) \right\vert^{2} \right\rangle
-\left\langle H\left( x\right) \right\rangle ^{2}\simeq \hbar ^{2}\Gamma
\frac{p_{i}^{2}}{2M^{2}}+\frac{\hbar ^{4}\Gamma ^{2}}{8M^{2}}  \label{30}
\end{equation}%
and as expected one regains the weak value energy time uncertainty relation
\begin{equation}
\left\langle \Delta t^{2}\left( x\right) \right\rangle \left\langle \Delta
H\left( x\right)\Delta H^*\left( x\right) \right\rangle \simeq \frac{\hbar
^{2}}{4}+\frac{\hbar ^{4}\Gamma }{16p_{i}^{2}}.  \label{31}
\end{equation}

The picture that emerges is thus that determining the momentum of a particle
at a point in space is analogous to determining the momentum of a particle
at a point in time. In the latter case, if one knows precisely the momentum,
then the position becomes fully indeterminate. In our case, pinpointing the
momentum is the same as pinpointing the energy and it is the time which
becomes indeterminate.

We now describe a second approach which also leads to determination of the
momentum at a given spatial point. Given the transition path time
distribution we have that the mean time it takes the particle to reach the
point $x$ is%
\begin{equation}
\left\langle t\left( x\right) \right\rangle =\int_{0}^{\infty }dttP\left(
t;x\right) .  \label{32}
\end{equation}%
The mean is well defined, since as noted, the long time tail of the
transition path time distribution goes as $t^{-3}$. We then consider the
mean time difference for the particle to reach two points in the scattering
direction $\left( x\right) $ which are chosen to be close to each other:%
\begin{equation}
\left\langle \delta t\left( x,\delta x\right) \right\rangle
=\int_{0}^{\infty }dtt\left[ P\left( t;x+\delta x/2\right) -P\left(
t;x-\delta x/2\right) \right] .  \label{33}
\end{equation}%
The momentum\ in the scattering direction is then by definition%
\begin{equation}
\bar{p}\left( x\right) =\lim_{\delta x\rightarrow 0}\frac{M\delta x}{\delta
t\left( x,\delta x\right) }=M\left( \frac{\partial \left\langle t\left(
x\right) \right\rangle }{\partial x}\right) ^{-1}.  \label{34}
\end{equation}%
With this protocol, which does not invoke a weak value, we determined a mean
momentum of the particle in the $x$ direction at the precise location $%
\left( x\right) $. Within the steepest descent approximation one readily
finds that
\begin{equation}
\bar{p}\left( x\right) ^{-1}\simeq \frac{2}{M}\int_{0}^{\infty
}dtP_{SD}\left( t;x\right) \left[ t\frac{\Gamma \left( x_{i}-x+\frac{p_{i}t}{%
M}\right) }{\left[ 1+\left( \frac{t\hbar \Gamma }{M}\right) ^{2}\right] }%
\right] \simeq \frac{1}{p_{i}}  \label{35}
\end{equation}

Using the square barrier model as above and performing all integrations
numerically exactly, we find that for $\Gamma =0.001$, the momentum $\bar{p}%
\left( x\right) =0.2502$. We thus find that also with this approach the mean
momentum at the post selected pointed $x$ is to a good approximation equal
to the mean incident momentum.

\section{Discussion}

The introduction of time averaging of weak values, using a transition path
time probability distribution leads to unexpected important results. We
showed that through time averaging it becomes possible to derive a rigorous
uncertainty relation for the product of the time averaged weak value of the
energy and the {\ (external)} time. Using the same formalism we also derived
a commutation relation for the time averaged commutator of the weak values
of the energy and the time. Within this formalism, the energy and the time
are analogous to the coordinate and momentum operators in quantum mechanics.
The two pairs obey the same uncertainty and commutation relations. This
result indicates that when considering time in quantum mechanics, one need
not construct a time operator. It is sufficient to consider time as a
parameter in the time dependent Schr{\"o}dinger equation, {\ or in the
terminology of Busch as an external time.} Equivalently it may be considered
to be its weak value as associated with the time dependent wavefunction at
the post selected point $x$.

The coordinate momentum uncertainty principle is derived for a measurement
of the two \textit{at the same value of the time interval, which is post
selected}. The energy time uncertainty relationship is derived for a fixed
value of the coordinate, which is post selected. {At} a given time it is
impossible to determine accurately both the momentum and the coordinate of a
particle. Similarly, at a given point in space, it is impossible to
determine accurately both the energy of the particle and the time at which
it will pass through the given point. On the other hand, just as it is
possible to determine accurately the position, or alternatively the momentum
of a particle at a fixed time, so it is possible to determine accurately the
momentum of a particle at a fixed position. We have demonstrated these
general relationships by considering explicitly the scattering of a particle
through a square well potential.

Finally, the localization of the position and the momentum is measurable
since the transition path time distribution is in principle measurable {%
as already noted in Section II. One places a screen at the
post-selected point $x$ and two synchronized clocks, one at the orifice of
the emerging particles and the other at the screen. In fact, single atom
time-of-flight experiments \cite{Fuhrmanek2010} have been implemented,
particles are released from a trap \cite{Du2015} at time zero, and the
arrival time at a detector screen is recorded. Similarly, a weak measurement
of the momentum at a post-selected point has been demonstrated
experimentally \cite{kocis11,flack14}. This means that if we prepare a
source of particles such that their mean momentum and spatial width {%
($1/\sqrt{\Gamma}$) are known, then we can predict with some
certainty the momentum of one of these particles when it arrives at the
post-selected (screen) point $x$. We cannot however, predict the arrival
time of the single particle with certainty.

\ack{ We thank Professor Ilya Rips
for insightful discussions. This work was supported by a grant from the
Israel Science Foundation and was partially supported by a grant with Ref.
FIS2014-52172-C2-1-P from the Ministerio de Economia y Competitividad
(Spain).}


\section*{References}

\begin{thebibliography}{99}
\bibitem{heisenberg27} Heisenberg W 1927 \textit{Z. Phys.} \textbf{43}
172-198.

\bibitem{ballentine98} Ballentine L E 2000 \textit{Quantum Mechanics: A
Modern Development}, (World Scientific, Singapore) Chap. 8.

\bibitem{busch08} Busch P 2008 \textit{Lect. Notes in Phys.} \textbf{734}
73-105.

\bibitem{ozawa03} Ozawa M 2003 \textit{Phys. Rev. A} \textbf{67} 042105.

\bibitem{lund10} Lund A P and Wiseman H M 2010 \textit{New J. Phys.} \textbf{%
12} 093011.

\bibitem{rozema12} Rozema L A, Darabi A, Mahler D H, Hayat A, Soudagar Y and
Steinberg A M 2012 \textit{Phys. Rev. Lett.} \textbf{109} 100404.

\bibitem{hilgevoord96} Hilgevoord J 1996 \textit{Am. J. Phys.} \textbf{64}
1451-6.

\bibitem{pauli33} Pauli W 1933 Die allgemeinen Prinzipien der
Wellenmechanik, in \textit{Handbuch der Physik}, edited by Geiger H, Scheel
K (Springer-Verlag, Berlin) Vol. 24.

\bibitem{pollak17c} Pollak E 2017 \textit{Phys. Rev. A} \textbf{95} 042108.

\bibitem{aharonov88} Aharonov Y, Albert D Z and Vaidman L 1988 \textit{Phys.
Rev. Lett.} \textbf{60} 1351-4.

\bibitem{wiseman07} Wiseman H M 2007 \textit{New J. Phys.} \textbf{9} 165.

\bibitem{mitchison07} Mitchison G, Josza R and Popescu S 2007 \textit{Phys.
Rev. A} \textbf{76} 062105.

\bibitem{tamir13} Tamir B and Cohen E 2013 \textit{Quanta} \textbf{2} 7-17.

\bibitem{petersen17} Petersen J and Pollak E 2017 \textit{J. Phys. Chem.
Lett.} \textbf{8} 4017-22.

\bibitem{Fuhrmanek2010} Fuhrmanek A, Lance A M, Tuchendler C, Grangier P,
Sortais Y R and Browaeys A 2010 \textit{New J. Phys.} \textbf{12} 053028.

\bibitem{Du2015} Du J-J, Li W-F, Wen R-J, Li G and Zhang T C 2015 \textit{%
Laser Phys. Lett.} \textbf{12} 065501.

\bibitem{cohentannoudji77} Cohen-Tannoudji C, Diu B and Laloe F 1977 \textit{%
Quantum Mechanics} (Wiley and Sons, New York) Vol. 1, pp 286-287.

\bibitem{muga08a} Muga G 2008 \textit{Lect. Notes in Phys.} \textbf{734}
31-72.

\bibitem{kocis11} Kocis S, Braverman B, Ravets S, Stevens M J, Mirin R P,
Shalm L K and Steinberg M A 2011 \textit{Science} \textbf{332} 1170-3.

\bibitem{flack14} Flack R and Hiley B J 2014 \textit{J. Phys. Conf. Ser.}
\textbf{504} 012016.
\end{thebibliography}
\end{document}